\documentclass{article}

\usepackage{amssymb,authblk,amsmath}
\usepackage{amsthm} 
\usepackage{graphicx}

\newtheorem{theorem}{Theorem}
\newtheorem{proposition}{Proposition}
\newtheorem{corollary}[theorem]{Corollary}

\begin{document}

\title{Perfect state transfer in two dimensions and the bivariate dual-Hahn polynomials}


\author{Hiroshi Miki}
\affil{Meteorological College, Asahi-Cho Kashiwa, 277 0852, Japan \thanks{hmiki@mc-jma.go.jp }}

\author{Satoshi Tsujimoto}
\affil{Department of Applied Mathematics and Physics, Graduate School of Informatics, Kyoto University, Sakyo-Ku Kyoto, 606 8501, Japan}

\author{Luc Vinet}
\affil{Centre de recherches math\'{e}matiques, Universit\'{e} de Montr\'{e}al, \\
PO Box 6128, Centre-ville Station, Montr\'{e}al (Qu\'{e}bec), H3C 3J7, Canada }
\affil{Institut de valorisation des donn\'{e}es (IVADO), Montr\`{e}al (Qu\'{e}bec), H2S 3H1, Canada}



\maketitle

\begin{abstract}%
A new solvable two-dimensional spin lattice model defined on a regular grid of triangular shape is proposed.
The hopping amplitudes between sites are related to recurrence coefficients of certain bivariate dual-Hahn polynomials. For a specific choice of the parameters, perfect state transfer and fractional revival are shown to take place. 
\end{abstract}

\section{Introduction}

Transporting quantum states between two sites is an important task in quantum information processing \cite{Kay1}.
Especially desirable is quantum state transfer realized with probability 1.
This is called perfect state transfer (PST).
Multi-site fractional revival (FR) is also useful for quantum entanglement generation \cite{Genest}.
Such protocols can be engineered in spin lattices \cite{Bose}. Recall that the single excitation dynamics of 1-dimensional $X\!X$ spin lattices with nearest-neighbor (inhomogeneous) couplings can be solved using orthogonal polynomials (OPs) \cite{Vinet2}.
It has been shown analytically that PST can be observed in the model where the corresponding wave functions are given in terms of Krawtchouk polynomials \cite{Christandl1}.
Building on this paradigm, a multi-dimensional extension has been proposed using multivariate Krawtchouk polynomials \cite{Miki1,Miki2,Miki3,Post1} and in some cases PST and FR were found to occur simultaneously.\par
This paper aims to provide an extension of these two-dimensional Krawtchouk spin lattice models.
To achieve that goal, we consider the dual-Hahn polynomials, which are a one-parameter generalization of the Krawtchouk polynomials.
It is known that PST happens in the one-dimensional spin lattice model associated with the dual-Hahn polynomials \cite{Chakrabarti}. 
With that in mind we will examine the possibility of using bivariate dual-Hahn polynomials to produce a two-dimensional model with interesting state transfer properties.
From the standard theory of OPs in several variables \cite{Dunkl}, it is apparent that many varieties of bivariate dual-Hahn polynomials are available; we shall focus on those of Tratnik type guided by the fact that it is that family of bivariate Krawtchouk polynomials that provided a two-dimensional model with PST \cite{Miki3}.
\par 
The paper is organized as follows. We first introduce the dual-Hahn polynomials and their bivariate extension of Tratnik type. We then construct a two-dimensional spin lattice model using those bivariate dual-Hahn polynomials.
We proceed to determine the transition amplitude for an excitation to move from one site to another in this model. The condition for PST is thereafter determined. Furthermore, the two-dimensional Krawtchouk spin lattice models are shown to be derived by taking a suitable limit. Conclusions will close the report.

\section{Bivariate dual-Hahn polynomials}
In this section, we review the basic properties of the univariate and bivariate dual-Hahn polynomials.
The univariate dual-Hahn polynomials are defined as follows \cite{Koekoek}:
\begin{align}\label{dual-hahn}
\begin{split}
d_n(x)&=d_n(x;\alpha ,\delta  ,\gamma )=(\alpha +1,\gamma+1)_n\cdot \mbox{}_3F_2 \left( \begin{matrix} -n,-x,x+\gamma +\delta +1 \\ \alpha+1 , \gamma +1\end{matrix};1\right)\\
&=(\alpha+1,\gamma+1)_n \sum_{l=0}^{\infty } \frac{(-n,-x,x+\gamma+\delta+1)_l}{(\alpha+1,\gamma+1)_l},
\end{split}
\end{align}
with 
\begin{align}
\begin{split}
&(a_1,a_2,\ldots ,a_m)_n=(a_1)_n(a_2)_n\ldots (a_m)_n,\\
&(a)_n=\begin{cases} a(a+1)\ldots (a+n-1)& (n=1,2,\ldots )\\
1&(n=0) 
\end{cases}
\end{split}
\end{align}
standard Pochhammer symbols.
Note that interchanging the role of $x$ and $n$ in \eqref{dual-hahn} gives the Hahn polnyomials \cite{Koekoek} and $d_n(x)$ is thus a polynomial of degree $n$ in the variable $x(x+\gamma +\delta +1)$.
When $\alpha +1 =-N$ and $\delta <-N, \gamma <-N$ (or $\delta >-1,\gamma >-1$), the orthogonality relation of the dual-Hahn polynomials is
\begin{equation}
\sum_{x=0}^N w_x(\alpha,\delta ,\gamma) d_n(x)d_m(x)=h_{n}(\alpha,\delta ,\gamma) \delta_{mn},
\end{equation}
where $\delta_{mn}$ is the Kronecker delta and 
\begin{align}
\begin{split}
w_x(\alpha ,\delta ,\gamma )&=\frac{(\gamma+\delta +1,\gamma/2+\delta/2+3/2,\alpha+1,\gamma+1)_x}{(\gamma/2+\delta/2+1/2,\gamma+\delta-\alpha+1,\delta+1)_xx!}(-1)^x\\
&=\frac{(\gamma+\delta +1,\gamma/2+\delta/2+3/2,-N,\gamma+1)_x}{(\gamma/2+\delta/2+1/2,\gamma+\delta+N+2,\delta+1)_xx!}(-1)^x,\\
\frac{h_n(\alpha,\delta,\gamma )}{n!}&=\frac{h_n}{n!}=(\alpha+1,\gamma+1,\alpha -\delta +1)_n\frac{\Gamma(\gamma +\delta -\alpha +1)\Gamma(\delta+1)}{\Gamma (\gamma +\delta +2)\Gamma(\delta -\alpha )}\\
&=(-N,\gamma+1,-N-\delta)_n\frac{(\gamma +\delta +2)_N}{(\delta+1)_N}
\end{split}
\end{align}
Being orthogonal and owing to Favard's theorem, the orthonormal dual-Hahn polynomials $\tilde{d}_n(x)=h_n^{-\frac{1}{2}}d_n(x)$ satisfy the three term recurrence relation which reads
\begin{align}
\begin{split}
&-x(x+\delta +\gamma +1 )\tilde{d}_n(x)\\
&=\sqrt{(n+1)(n+\gamma +1)(N-n)(\delta +N-n)}\tilde{d}_{n+1}(x)\\
&-\{ (n+\gamma+1)(N-n)+n (\delta +N-n+1)\} \tilde{d}_n(x)\\
&+\sqrt{n(n+\gamma )(N-n+1)(\delta +1+N-n)}\tilde{d}_{n-1}(x).
\end{split}
\end{align} 

The bivariate dual-Hahn polynomials of Tratnik type are given as the following product of univariate dual-Hahn polynomials \cite{Tratnik1}:
\begin{align}\label{2var-hahn}
\begin{split}
D_{m,n}=D_{m,n}(x,y)&=d_m(x;b+y-1,a+y,-y-1)\\
&\cdot d_n(y-m,m+c+N-1,m+b+N,m-N-1)
\end{split}
\end{align}
for $0\le x\le y\le N$ and $0\le m+n\le N$.
It should be remarked that the notations are different from the ones in \cite{Tratnik1} and that the polynomials \eqref{2var-hahn} coincide with those in \cite{Tratnik1} if we take $a=\alpha_1,b=\alpha_1+\alpha_2,c=\alpha_1+\alpha_2+\alpha_3$. These polynomials can be obtained as a limit of the bivariate Racah polynomials \cite{Geronimo}. Their orthogonality relation is given by
\begin{align}\label{2var:orthogonality}
\begin{split}
\sum_{y=0}^N\sum_{x=0}^yD_{m_1,n_1}(x,y)D_{m_2,n_2}(x,y)w_{x,y}=r_{m_1,n_1}\delta_{m_1,m_2}\delta_{n_1,n_2},
\end{split}
\end{align}
where 
\begin{align}
\begin{split}
w_{x,y}=&\frac{(-1)^x}{x!(y-x)!}\cdot \frac{\Gamma (b-a+y-x) \Gamma (b+y+x)}{\Gamma (a+1+y+x)}\cdot \frac{(a,a/2+1)_x}{(a/2)_x},\\
&\cdot \frac{(b/2+1,c+N,-N)_y}{(b/2,c-b-N+1,b+N+1)_y}\\
r_{m,n}=&m!n!(-N,c+N)_{m+n}\frac{\Gamma (b-a+m)\Gamma (c-b+n)\Gamma (b+N+1)}{(-1)^Nab\Gamma (a)\Gamma (c-b+N)}.
\end{split}
\end{align}
We see from its expression that the weight function $w_{x,y}$ is positive if 
\begin{equation}\label{cond}
c>0,\quad a-b>N,\quad b-c>N;
\end{equation}
we shall assume that this holds in what follows.
Taking into account that
\begin{align}\label{rel1}
\begin{split}
&\frac{w_{y-m}(m+c+N-1,m+b+N,m-N-1)}{h_n(m+c+N-1,m+b+N,m-N-1)}\\
&=\frac{h_m(b+y-1,a+y,-y-1)w_{x,y}}{w_x(b+y-1,a+y,-y-1)r_{m,n}},
\end{split}
\end{align}
the orthogonality relation \eqref{2var:orthogonality} can be verified by applying the orthogonality relation of the univariate dual-Hahn polynomials twice. From \eqref{rel1}, we find the following relation:
\begin{equation}\label{rel2}
\frac{r_{i,j}}{r_{i,l}}=\frac{h_j(i+c+N-1,i+c+N,i-N-1)}{h_l(i+c+N-1,i+c+N,i-N-1)},
\end{equation}
which will prove useful later.
The bivariate dual-Hahn polynomials possess the property:
\begin{proposition}
For $0\le x\le y\le N$, $\{ D_{m,n}(x,y)\}_{0\le m+n\le N}$ obeys the following recurrence relations:
\begin{align}
\begin{split}
&x(x+a)D_{m,n}=D_{m+1,n}+D_{m,n+1}\\
&+\{ N(N+c)-m(c+b-a-1)-n(2c-b-1)\\
&-2m^2-2mn-2n^2\}D_{m,n}\\
&+m(a-b+1-m)(N+1-m-n)(c+N+m+n-1)D_{m-1,n}\\
&+n(b-c+1-n)(N+1-m-n)(c+N+m+n-1)D_{m,n-1}\\
&+m(a-b+1-m)D_{m-1,n+1}+n(b-c+1-n)D_{m+1,n-1},\\
&y(y+b)D_{m,n}=D_{m,n+1}\\
&+\{ N(N+c)-n(2c-b-1)-m(c-b)-2mn-2n^2\} D_{m,n}\\
&+n(b-c+1-n)(N+1-m-n)(c+N+m+n-1)D_{m,n-1}.
\end{split}
\end{align}
\end{proposition}
These relations can be obtained from limits of the recurrence relations of the bivariate Racah polynomials \cite{Geronimo}.
Subtracting the above relations pairwise, we have 
\begin{align}\label{2var-hahn:contiguous}
\begin{split}
&(x(x+a)-y(y+b))D_{m,n}\\
&=D_{m+1,n}+m(a-2b+1-2m)D_{m,n}\\
&+m(a-b+1-m)(N+1-m-n)(c+N+m+n-1)D_{m-1,n}\\
&+m(a-b+1-m)D_{m-1,n+1}+n(b-c+1-n)D_{m+1,n-1}.
\end{split}
\end{align}
For the orthonormal polynomials $\tilde{D}_{m,n}(x,y)=r_{m,n}^{-\frac{1}{2}}D_{m,n}(x,y)$, the relation \eqref{2var-hahn:contiguous} becomes
\begin{align}
\begin{split}
\lambda _{x,y}\tilde{D}_{m,n}(x,y)&=J_{m+1,n}\tilde{D}_{m+1,n}+J_{m,n}\tilde{D}_{m-1,n}+B_{m,n}\tilde{D}_{m,n}\\
&+L_{m,n}\tilde{D}_{m-1,n+1}+L_{m+1,n-1}\tilde{D}_{m+1,n-1}
\end{split}
\end{align}
with
\begin{align}\label{rec:coeff}
\begin{split}
\lambda_{x,y}&=x(x+a)-y(y+b),\\
J_{m,n}&=\sqrt{m(N+1-m-n)(a-b+1-m)(c+N+m+n+1)},\\
B_{m,n}&=m(a-2b+1-2m),\\
L_{m,n}&=\sqrt{m(n+1)(a-b-m+1)(b-c-n)}.
\end{split}
\end{align}

\section{The spin lattice model}
We consider in this section the dynamics of a spin lattice of triangular shape governed by the Schr\"{o}dinger equation
\begin{equation}
i\hbar \frac{d }{d t}\left| \psi (t)\right> =H \left| \psi (t)\right> 
\end{equation}
with the following Hamiltonian:
\begin{align}
\begin{split}\label{hamiltonian}
H=&\sum_{0\le i+j\le N} X_{i+1,j}\frac{\sigma_{i,j}^x \sigma_{i+1,j}^x+\sigma_{i,j}^y\sigma_{i+1,j}^y}{2}\\
&\quad \quad + Y_{i+1,j-1}\frac{\sigma_{i,j}^x \sigma_{i+1,j-1}^x+\sigma_{i,j}^y\sigma_{i+1,j-1}^y}{2}+ Z_{i,j} \frac{1+\sigma_{i,j}^z}{2}
\end{split}
\end{align}  
with $X_{0,j}=Y_{0,j}=0$ and $X_{i,j}=Y_{i,j}=0\quad (i+j>N)$.
The lattice sites are labelled by two integers $i,j$ between $0$ and $N$ such that $0\le i+j\le N$ and $\sigma_{i,j}^x,\sigma_{i,j}^y,\sigma_{i,j}^z$ are Pauli matrices acting on the site labelled by $(i,j)$.
The constants $X_{i,j}$ and $Y_{i,j}$ are coupling the sites $(i,j)$ and $(i-1,j)$ and the sites $(i,j)$ and $(i-1,j+1)$ respectively. The constants $B_{i,j}$ are the strengths of the magnetic fields at the sites $(i,j)$. In what follows, we scale the time variable $t$ so that $\hbar =1$.\par  
It is not difficult to see that the Hamiltonian \eqref{hamiltonian} is invariant under rotations about the $z$ axis:
\begin{equation}
\left[ H, \sum_{0\le i+j\le N} \sigma_{i,j}^z\right]=0,
\end{equation}
which implies that it preserves the total number of spins that are up (or down).
Let us denote the 1-excitation basis vectors by $\left| e_{i,j}\right)=E_{i,j}$ with $E_{i,j}$ the $(N+1)\times (N+1)$ matrix with $1$ in the $(i,j)$ entry and zeros everywhere else.
The action of $H$ on the subspace spanned by these 1-excitation basis vectors is given by
\begin{align}\label{action_Hamiltonian}
\begin{split}
H\left| e_{i,j}\right)
&=X_{i+1,j}\left| e_{i+1,j}\right)+X_{i,j}\left| e_{i-1,j}\right)\\
&+Y_{i+1,j-1}\left| e_{i+1,j-1}\right)+Y_{i,j}\left| e_{i-1,j+1}\right)\\
&+Z_{i,j}\left| e_{i,j}\right).
\end{split}
\end{align}
From now on, we shall restrict $H$ to the subspace spanned by $\{ \left| e_{i,j}\right) \}$, which implies that the Hamiltonian $H$ can be identified with the representation matrix determined by \eqref{action_Hamiltonian}. We consider the eigenvalue problem:
\begin{equation}\label{evp}
H\left| x,y\right> = \lambda_{x,y} \left| x,y\right>,
\end{equation}
where $\left| x,y\right>$ is the eigenstate of $H$ with eigenvalues $\lambda_{x,y}$.
Consider the expansion of $\left| x,y\right> $ over the basis vectors $\left| e_{i,j}\right)$:
\begin{equation}\label{expansion}
\left| x,y\right> = \sum_{0\le i+j\le N} W_{i,j}(x,y)\left| e_{i,j}\right). 
\end{equation}
From \eqref{evp}, we see that the coefficients $W_{i,j}(x,y)$ satisfy the contiguous relations:
\begin{align}\label{contiguous}
\begin{split}
\lambda_{x,y}W_{i,j}&=X_{i+1,j}W_{i+1,j}+X_{i,j}W_{i,j}+Z_{i,j}W_{i,j}\\
&+Y_{i,j}W_{i-1,j+1}+Y_{i+1,j-1}W_{i+1,j-1}.
\end{split}
\end{align} 
While not solvable in general the eigenvalue problem \eqref{evp} (or the contiguous relation \eqref{contiguous}) can be solved for certain choices of the couplings and Zeeman terms.
One such choice is
\begin{align}\label{par-kraw0}
\begin{split}
X_{i,j}&=\sqrt{2i(N+1-i-j)},\quad 
Y_{i,j}=\sqrt{2i(j+1)},\quad 
Z_{i,j}=0.
\end{split}
\end{align}
This parametrization comes from the bivariate Krawtchouk polynomials and remarkably perfect state transfer takes place in this case \cite{Miki2}.\par
Comparing \eqref{contiguous} with relation \eqref{2var-hahn:contiguous}, we are led to explore the following new choice of couplings and magnetic fields.
\begin{proposition}
Set
\begin{equation}
X_{i,j}=J_{i,j},\quad Y_{i,j}=L_{i,j},\quad Z_{i,j}=B_{i,j},
\end{equation} 
using the coefficients of the recurrence relations of bivariate dual-Hahn polynomials \eqref{rec:coeff}. Then the solution to the eigenvalue problems \eqref{evp} is explicitly given in terms of bivariate dual-Hahn polynomials:
\begin{align}
\begin{split}
\lambda_{x,y}=x(x+a)-y(y+b),\quad W_{i,j}(x,y)=\sqrt{w_{x,y}}\tilde{D}_{i,j}(x,y).
\end{split}
\end{align}
\end{proposition}
We shall now focus on the 1-excitation dynamics in the spin lattice thus defined.
As a companion to \eqref{expansion}, in view of \eqref{2var:orthogonality}, we have the inverse expansion
\begin{equation}\label{inv-expansion}
\left| e_{i,j}\right)=\sum_{0\le x\le y\le N}W_{i,j}(x,y)\left| x,y\right>.
\end{equation}
The transition amplitude $f_{(i,j),(k,l)}(T)$ for an excitation initially at the site $(i,j)$ to be found at the site $(k,l)$ after some time $T$ is determined by 
\begin{equation}
f_{(i,j),(k,l)}(T)=\left( e_{k,l}| \psi (T)\right>
\end{equation}
with the initial value $\left| \psi (0)\right> =\left| e_{i,j}\right)$. Therefore, we have
\begin{equation}
f_{(i,j),(k,l)}(T)=\left( e_{k,l}|\exp (-iTH)|e_{i,j}\right).
\end{equation}
With the help of \eqref{expansion} and \eqref{inv-expansion}, the transition amplitude $f_{(i,j),(k,l)}(T)$ is given by
\begin{align}
\begin{split}
f_{(i,j),(k,l)}(T)
&=\sum_{0\le x\le y\le N}W_{i,j}(x,y)W_{k,l}(x,y)e^{-iT\lambda_{x,y}}\\
&=\frac{1}{\sqrt{r_{i,j}r_{k,l}}}\sum_{0\le x\le y\le N}w_{x,y}D_{i,j}(x,y)D_{k,l}(x,y)e^{-iT(x(x+a)-y(y+b))}.
\end{split}
\end{align}
We can calculate this transition amplitude by using the explicit expression for the bivariate dual-Hahn polynomials.
In general, this becomes a sum of hypergeometric series and cannot be simplified further. However, if we choose the parameters $a,b$ and the time $T$ so that
\begin{equation}\label{pst_cond1}
e^{-iTx(x+a)}=1,\quad e^{-iTy(y+b)}=(-1)^y,
\end{equation} 
the amplitude can be reduced to
\begin{align}\label{amplitude1}
\begin{split}
&f_{(i,j),(k,l)}(T)\\
&=\frac{1}{\sqrt{r_{i,j}r_{k,l}}}\sum_{y=0}^N\sum_{x=0}^yw_{x,y}(-1)^yd_i(x;b+y-1,a+y,-y-1)\\
&\qquad \qquad \cdot d_j(y-i,i+c+N-1,i+b+N,i-N-1)\\
&\qquad \qquad \cdot  d_k(x;b+y-1,a+y,-y-1)\\
&\qquad \qquad \cdot d_l(y-k,k+c+N-1,k+b+N,k-N-1)\\
&=\frac{r_{i,j}\delta_{i,k}}{\sqrt{r_{i,j}r_{i,l}}}\sum_{y=i}^N \frac{w_{y-i}(i+c+N-1,i+c+N,-(N-i)-1)}{h_j(i+c+N-1,i+c+N,i-N-1)}(-1)^y\\
&\qquad \qquad \cdot d_j(y-i,i+c+N-1,i+b+N,i-N-1)\\
&\qquad \qquad \cdot d_l(y-i,i+c+N-1,i+b+N,i-N-1)\\
&=\delta_{i,k}\sum_{y=i}^N w_{y-i}(i+c+N-1,i+c+N,-(N-i)-1)(-1)^y\\
&\qquad \qquad \cdot \frac{d_j(y-i,i+c+N-1,i+b+N,i-N-1)}{\sqrt{h_j(i+c+N-1,i+c+N,i-N-1)}}\\
&\qquad \qquad \cdot \frac{d_l(y-i,i+c+N-1,i+b+N,i-N-1)}{\sqrt{h_l(i+c+N-1,i+c+N,i-N-1)}},
\end{split}
\end{align}
where we have used \eqref{rel2}.
It is not difficult to see that the transition amplitude $f_{(i,k),(i,l)}(T)$ can be put in correspondance with the one for the 1-dimensional spin lattice model associated with univariate dual-Hahn polynomials \cite[Sec. 4.1]{Chakrabarti}.
Therefore, according to the result of the 1-dimensional case, we have in particular 
\begin{align}
\begin{split}
f_{(0,0),(0,N)}(T)=\frac{\sqrt{(c+N,b+1-c-N)_N}}{\left( \frac{b+1}{2}\right)_N}.
\end{split}
\end{align}
Remarkably, if we set 
\begin{equation}\label{pst_cond}
b=2c+2N-1,
\end{equation}
it is easy to see that
\begin{equation}\label{pst}
|f_{(0,0),(0,N)}|=1,
\end{equation}
which means that perfect state transfer from $(0,0)$ to $(0,N)$ takes place.
From the condition \eqref{pst_cond1}, one can see that $a,b$ are rational and we now set
\begin{equation}
a=\frac{p'}{k'},\quad b=\frac{q'}{k'},\quad (k',p',q'\in \mathbb{N}).
\end{equation}
Then at time $T=k'\pi$, the condition \eqref{pst_cond1} can be rewritten as 
\begin{subequations}
\begin{align}
&x(k'x+p')=2l', \label{pst_cond3a}\\
&y(k'y+q')=\begin{cases} 2m' & (y:\textrm{even}) \\ 2n'-1 & (y:\textrm{odd})\end{cases} \quad (l',m',n'\in \mathbb{N}). \label{pst_cond3b}
\end{align}
\end{subequations}
Since $x$ and $y$ are integers, it is not difficult to see that the relation \eqref{pst_cond3a} holds when both $k'$ and $p'$ are even or odd. The relation \eqref{pst_cond3b} holds when $k'$ is even and $q'$ is odd or when $k'$ is odd and $q'$ is even.
As a consequence, we have.
\begin{theorem}\label{thm2}
If we take
\begin{equation}\label{pst_par1}
a= \frac{2p-1}{2k-1},\quad b=\frac{2q}{2k-1},\quad c=\frac{b}{2}-N+\frac{1}{2}\quad (k,p,q\in \mathbb{N})
\end{equation}
and assume \eqref{cond} satisfied, the relation \eqref{pst_cond1} is verified at $T=(2k-1)\pi $
and perfect state transfer between $(0,0)$ and $(0,N)$ occurs.
Furthermore, if we take 
\begin{equation}\label{pst_par2}
a= \frac{p}{k},\quad b=\frac{2q-1}{2k},\quad c=\frac{b}{2}-N+\frac{1}{2}\quad (k,p,q\in \mathbb{N}),
\end{equation}
relation \eqref{pst_cond1} is verified at $T=2k\pi $
and perfect state transfer is also observed between $(0,0)$ and $(0,N)$.
\end{theorem} 

\begin{figure}
  \begin{center}
  \includegraphics[width=10cm]{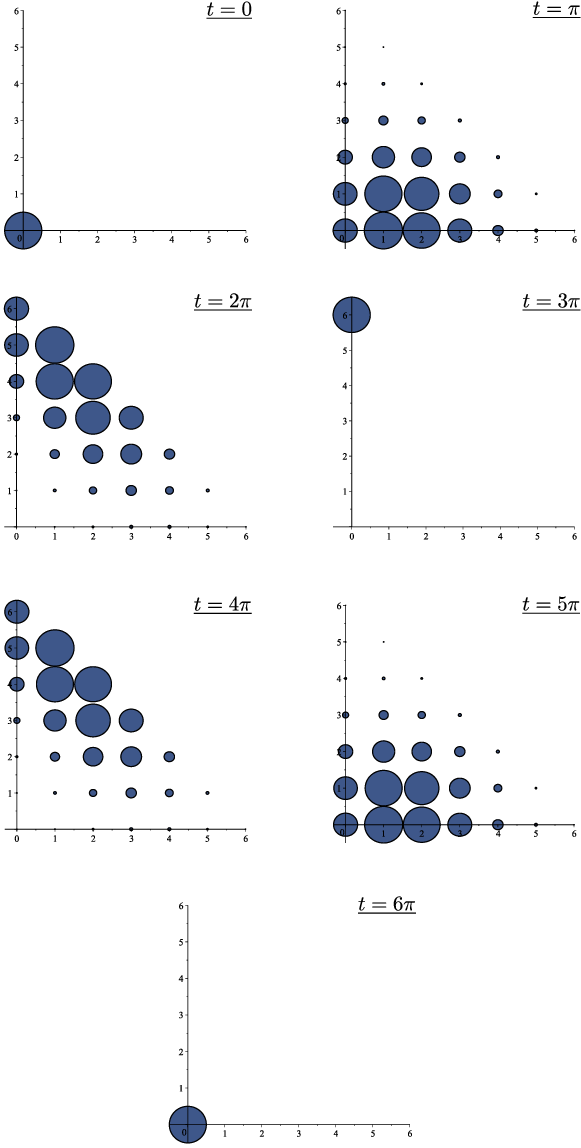}
    \caption{The plot of the modulus of the transition amplitude $|f_{(0,0),(i,j)}(t)|$ when $a=\frac{53}{3},b=\frac{34}{3},c=\frac{1}{6}$ and $N=6$. The areas of the circles
are proportional to $|f_{(0,0),(i,j)}(t)|$ at the given lattice point $(i,j)$. PST from $(0,0)$ to $(0,6)$ occurs at $t=3\pi $.}
    \label{fig:1}
  \end{center}
\end{figure}

Taking \eqref{cond} into account, the condition \eqref{pst_par1} is satisfied for instance when 
\begin{equation}
N=6,\quad a=\frac{53}{3},\quad b=\frac{34}{3},\quad c=\frac{1}{6},
\end{equation}
which amounts to 
\begin{align}
\begin{split}
X_{i,j}&=\frac{\sqrt{2i(7-i-j)(22-3i)(43+6i+6j)}}{6},\\
Y_{i,j}&=\frac{\sqrt{2i(j+1)(22-3i)(67-6j)}}{6},\\
Z_{i,j}&=-2i(2+i).
\end{split}
\end{align}
In this case, PST from $(0,0)$ to $(0,6)$ can be observed at $t=T=3\pi$ and this is illustrated in Fig. \ref{fig:1}.
It should be remarked that the image is symmetric with respect to $t=T=3\pi$, which turns out that the state transfer is periodic with the periodicity $2T=6\pi$. This can be explained by the fact that $e^{-2iTx(x+a)}=e^{-i2Ty(y+b)}=1$ and the orthogonality relation \eqref{2var:orthogonality}.
\par 
In addition to PST, we can observe fractional revival in particular cases.

\begin{figure}
    \begin{center}
  \includegraphics[width=12cm]{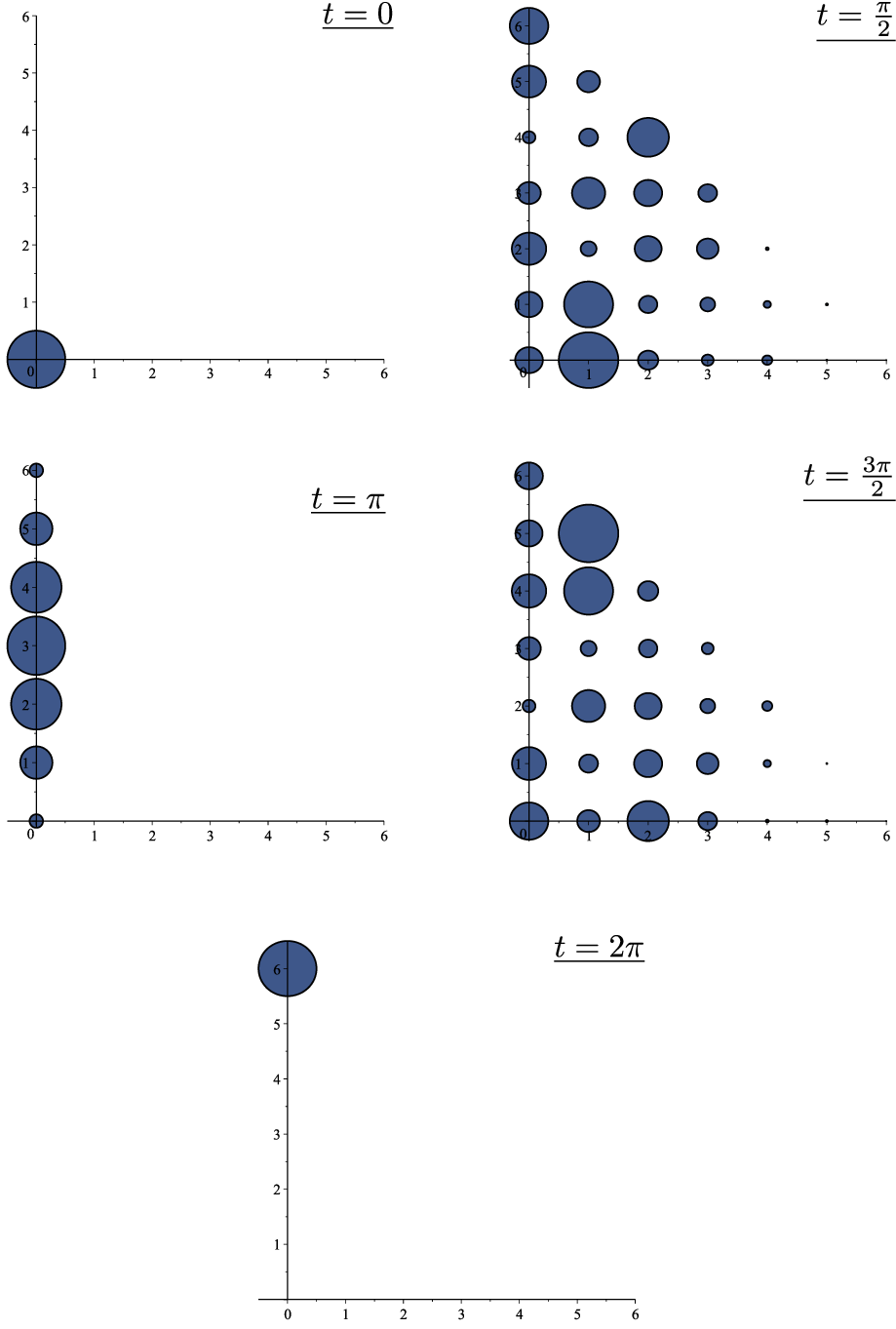}
    \caption{The plot of the modulus of the transition amplitude $|f_{(0,0),(i,j)}(t)|$ when $a=19,b=\frac{23}{2},c=\frac{1}{4}$ and $N=6$. The areas of the circles
are proportional to $|f_{(0,0),(i,j)}(t)|$ at the given lattice point $(i,j)$. PST from $(0,0)$ to $(0,6)$ occurs at $t=2\pi $. Fractional revival also occurs at $t=\pi $.}
    \label{fig:2}
  \end{center}
\end{figure}

\begin{corollary}
Assume \eqref{pst_par2} and $p$ to be odd, which amounts to 
\begin{equation}
e^{-i\frac{T}{2}x(x+a)}=1,\quad e^{-i\frac{T}{2}y(y+b)}\ne (-1)^y.
\end{equation}  
Then fractional revival from $(0,0)$ to the sites  $(0,j)$ takes place at $\frac{T}{2}$ that is, at times equal to $k\pi $.
\end{corollary}
The condition \eqref{pst_par2} is satisfied for instance when
\begin{equation}
N=6,\quad a=19,\quad b=\frac{23}{2},\quad c=\frac{1}{4},
\end{equation}
which amounts to 
\begin{align}
\begin{split}
X_{i,j}&=\frac{\sqrt{2i(7-i-j)(17-2i)(29+4i+4j)}}{4},\\
Y_{i,j}&=\frac{\sqrt{2i(j+1)(17-2i)(45-4j)}}{4},\\
Z_{i,j}&=-i(3+2i).
\end{split}
\end{align}
In this case, PST from $(0,0)$ to $(0,6)$ can be observed at $t=2\pi $ and fractional revival can be found at $t=\pi$. This is illustrated in Fig. \ref{fig:2} 

As for PST in one-dimension \cite{Kay1}, a form of mirror-symmetry is also observed here. Indeed we find that PST takes place between sites that are mirror images with respect to a horizontal line passing through the vertical of the lattice on which they are located. In other words we find that

\begin{figure}
   \begin{center}
  \includegraphics[width=12cm]{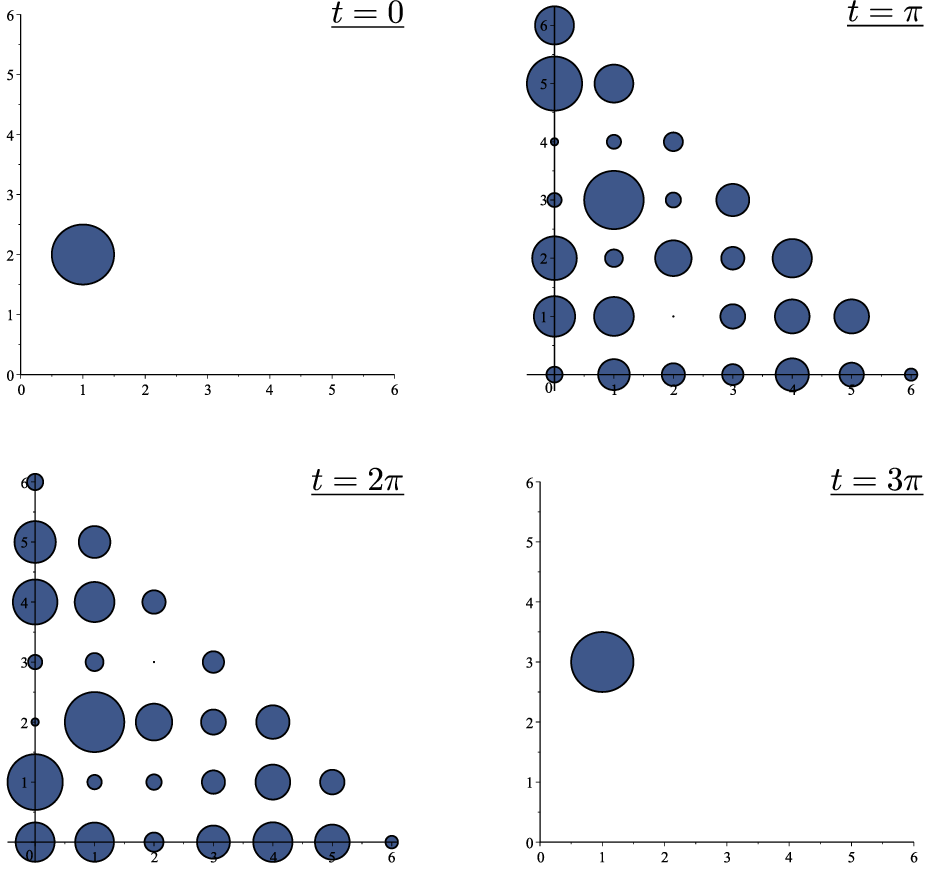}
    \caption{The plot of the modulus of the transition amplitude $|f_{(1,2),(i,j)}(t)|$ when $a=\frac{53}{3},b=\frac{34}{3},c=\frac{1}{6}$ and $N=6$. The areas of the circles
are proportional to $|f_{(1,2),(i,j)}(t)|$ at the given lattice point $(i,j)$. PST from $(1,2)$ to $(1,3)$ occurs at $t=3\pi $.}
    \label{fig:3}
  \end{center}
\end{figure}

\begin{corollary}
With the same assumptions as Proposition \ref{thm2}, we also have other perfect state transfer situations (with different initial sites) over the same period of time $T$:
\begin{equation}
|f_{(i,j),(i,N-i-j)}(T)|=1\quad (0\le i+j\le N).
\end{equation}
\end{corollary}
This is also illustrated in Fig. \ref{fig:3}.

\section{Degeneration to the 2-dimensional Krawtchouk model}
It is known that the ordinary Krawtchouk polynomials
\begin{equation}
K_n^N(x;p)= \mbox{}_2F_1 \left( \begin{matrix} -n,-x \\ -N \end{matrix};\frac{1}{p}\right),\quad 0<p<1
\end{equation} 
can be obtained as a degeneration limit of the ordinary dual-Hahn polynomials
\begin{equation}
D_{n}(x;\alpha ,\beta ,\gamma )= \mbox{}_3F_2 \left( \begin{matrix} -n,-x,x+\gamma +\delta +1 \\ \alpha+1 , \gamma +1\end{matrix};1\right)
\end{equation}
as follows \cite{Koekoek}:
\begin{equation}
\lim_{t\to \infty }D_n(x;-N-1,(1-p)t,pt)=K_n^N(x;p).
\end{equation}
In addition, we have another limit formula which will be useful here:
\begin{equation}\label{limit1}
\lim_{t\to \infty }D_n(x;p_1t+k_1,p_2t+k_2,-N-1)=K_n^N\left( x; \frac{p_1}{p_2}\right)\quad (p_2,p_1,k_1,k_2\in \mathbb{R}).
\end{equation}
Applying \eqref{limit1}, we can obtain the bivariate Krawtchouk polynomials from the bivariate dual-Hahn polynomials \eqref{2var-hahn}. 
\begin{theorem}
Let 
\begin{equation}
a=At+o(t),\quad b=Bt+o(t),\quad c=Ct+o(t).
\end{equation}
Then the following limit formula holds:
\begin{equation}
\lim_{t\to \infty }t^{-m-n}D_{m,n}(x,N-y)\propto K_{n,m}^N\left( y,x;\frac{B-C}{B},\frac{C}{A}\right),
\end{equation}
where $K_{m,n}^N(u,v;p,q)$ are the bivariate Krawtchouk polynomials of Tratnik type \cite{Tratnik1} defined by
\begin{equation}
K_{m,n}^N(u,v;p,q)=\frac{(n-N)_m(u-N)_n}{(-N)_{m+n}}K_m^{N-n}\left( u;p\right) K_n^{N-u}\left( v;\frac{q}{1-p}\right)
\end{equation}
for $0\le u+v,m+n\le N$.
\end{theorem}
\begin{proof}
One can verify that  
\begin{equation}
D_n(x;a,b,-N-1)=\frac{(a-b+1)_n}{(a+1)_n}D_n(N-x,a-b,-b,-N-1).
\end{equation}
Using this relation, we can rewrite the bivariate dual-Hahn polynomials as follows:
\begin{align}
\begin{split}
D_{m,n}(x,N-y)&=(b+N-y,y-N)_m(m+c+N,m-N)_n\\
&\cdot D_n(N-y-m,m+c+N-1,m+b+N,m-N-1)\\
&\cdot D_m(x,b+N-y-1,a+N-y,y-N-1)\\
&=(c-b)_n(b-a)_m(m-N)_n(y-N)_m\\
&\cdot D_n(y,c-b-1,-m-b-N,m-N-1)\\
&\cdot D_m(N-y-x,b-a-1,y-N-a,y-N-1)
\end{split}
\end{align}
Therefore, from \eqref{limit1}, one can see that
\begin{align}
\begin{split}
&\lim_{t\to \infty }t^{-m-n}D_{m,n}(x,N-y)\\
&=\lim_{t\to \infty }\left\{  t^{-m-n}(m-N)_n(y-N)_m ((C-B)t+o(t))_n((A-B)t+o(t))_m\right.\\
&\quad \cdot D_n(y,-t(B-C)+o(t),-tB+o(t),m-N-1)\\
&\left.\quad \cdot D_m(N-y-x,-t(A-B)+o(t),-tA+o(t),y-N-1) \right\}\\
&=(C-B)^n(B-A)^m (m-N)_n(y-N)_mK_n^{N-m}\left( y;\frac{B-C}{B}\right)K_m^{N-y}\left( N-x-y;\frac{A-B}{A}\right)\\
&=(C-B)^n(B-A)^m(m-N)_n(y-N)_mK_n^{N-m}\left( y;\frac{B-C}{B}\right)K_m^{N-y}\left( N-x-y;\frac{A-B}{A}\right)\\
&=(C-B)^nB^m (m-N)_n(y-N)_mK_n^{N-m}\left( y;\frac{B-C}{B}\right)K_m^{N-y}\left( x;\frac{B}{A}\right)\\
&\propto K_{n,m}^N\left( y,x;\frac{B-C}{B},\frac{C}{A}\right),
\end{split}
\end{align}
where we have used 
\begin{equation}
K_n^N(x;1-p)=\left( \frac{p}{p-1}\right)^n K_n^N(N-x;p).
\end{equation}
This completes the proof.
\end{proof}
Observing \eqref{rec:coeff}, the parameters of the Hamiltonian \eqref{hamiltonian} corresponding to the bivariate Krawtchouk polynomials are easily calculated as follows
\begin{align}\label{par-kraw}
\begin{split}
X_{i,j}&=\lim_{t\to \infty }\frac{J_{i,j}}{t}=\sqrt{(A-B)Ci(N+1-i-j)},\\
Y_{i,j}&=\lim_{t\to \infty }\frac{L_{i,j}}{t}=\sqrt{(A-B)(B-C)i(j+1)},\\
Z_{i,j}&=\lim_{t\to \infty }\frac{B_{i,j}}{t}=(A-2B)i
\end{split}
\end{align}
and the corresponding eigenvalues are explicitly given by
\begin{equation}
\lambda_{x,y}=\lim_{t\to \infty }\frac{x(x+a)-y(y+b)}{t}=Ax-By,\quad 0\le x\le y\le N.
\end{equation} 
It should be remarked that in the previous section we can observe PST with the condition \eqref{pst_cond}, which amounts to 
\begin{equation}
B=2C.
\end{equation} 
Then we have the following result about PST in the bivariate Krawtchouk model.
\begin{corollary}
The Hamiltonian $\eqref{hamiltonian}$ with \eqref{par-kraw} and $B=2C$ can be diagonalized by the bivariate Krawtchouk polynomials: 
\begin{equation}
W_{i,j}(x,y)=\sqrt{\frac{\hat{w}_{x,N-y}}{\hat{r}_{n,m}}}K_{n,m}^N\left( N-y,x;p,q\right),
\end{equation}
where $p=\frac{1}{2},q=\frac{C}{A}$ and
\begin{align}
\begin{split}
&\hat{w}_{x,y}=\binom{N}{x,y}p^xq^{y}(1-p-q)^{N-x-y},\\
&\hat{r}_{m,n}=\frac{(1-p-q)^{m+n}}{\binom{N}{m,n}\left( \frac{p(1-p-q)}{1-p}\right)^m\left( \frac{q}{1-p}\right)^n}.
\end{split}
\end{align}
Furthermore, when 
\begin{equation}
q=\frac{2l-1}{4k}\quad (k,l\in \mathbb{N}),
\end{equation}
there exist some time $T$ such that $e^{-iATx}=1,e^{-iBTy}=(-1)^y$ and at that time $T$ the following relation holds:
\begin{equation}
|f_{(0,0),(0,N)}(T)|=1.
\end{equation}
\end{corollary}
If we set $B=2C$ and $A=4C$, the coefficients \eqref{par-kraw} are essentially equal to \eqref{par-kraw0}, which is treated in \cite{Miki2}.
Thus the result here is its generalization.
\section{Concluding Remarks}
To sum up, we have introduced a new two-dimensional spin lattice model associated with the bivariate dual-Hahn polynomials.
Since these polynomials are a one-parameter generalization of the bivariate Krawtchouk polynomials, the model presented here is an extension of the spin lattice discussed in \cite{Miki2,Miki3}.
Furthermore, for specific parameters, we have found that perfect state transfer (PST) happens in the model.
This fact is based on the observation that some transition amplitudes can be identified with those of the one-dimensional spin lattice associated with univariate dual-Hahn polynomials.
This suggests that it could be possible to relate formally our two-dimensional model to the chain based on the univariate dual-Hahn polynomials (for the specific choice of parameters). We have not succeeded to do this however.
By introducing a degeneration limit from the bivariate dual-Hahn polynomials to the Krawtchouk polynomials, we have obtained the two-dimensional Krawtchouk models with PST including a model in \cite{Miki2} as a special case.  
\par
In \cite{Miki2}, the model based on the bivariate Krawtchouk polynomials was related to PST on graphs of the ordered Hamming scheme. This association scheme generalizes the Hamming one to which the univariate Krawtchouk polynomials are associated. It would be quite interesting to lift the results obtained here to graphs of some scheme \cite{Bernard,Chan,Coutinho}. The dual-Hahn polynomials are attached to the Johnson scheme \cite{Stanton}. It should be noted however that while PST occurs in a spin chain based on the dual-Hahn polynomials, no lifts are possible to unweighted graphs of the Johnson scheme \cite{Ahmadi,Vinet} owing to parameter incompatibility. Moreover to our knowledge no analogue of the ordered Hamming scheme to which multivariate dual-Hahn would be connected has been designed. We will leave these interesting questions to future considerations.

\section*{Acknowledgement}
The authors would like to thank an anonymous referee for reading the manuscript carefully and the helpful advices which improved the manuscript.
The research of HM and ST is supported by JSPS KAKENHI (Grant Numbers 21H04073 and 19H01792 respectively) and that of LV by a discovery grant of the Natural Sciences and Engineering Research Council (NSERC) of Canada. 

\let\doi\relax

\end{document}